\documentstyle[psfig]{mn}

\def\deg{\ifmmode^{\circ}\;\else$^{\circ}\;$\fi}

\begin{document}
\title[A search for giant pulses in Vela-like pulsars]
      {A search for giant pulses in Vela-like pulsars}

\author[Johnston \& Romani]{Simon~Johnston$^1$ \&
Roger~W.~Romani$^{2}$\\
$^1$School of Physics, University of Sydney, NSW 2006, Australia\\
$^2$Dept. of Physics, Stanford University, Stanford CA  94305-4060, USA}

\maketitle
\begin{abstract}
We have carried out a survey for `giant pulses' in 6 young, Vela-like
pulsars. In no cases did we find single pulses with flux densities
more than 10 times the mean flux density. However, in 
PSR B1706--44 we have detected giant micro-pulses very
similar to those seen in the Vela pulsar. In PSR B1706--44 these giant
micro-pulses appear on the trailing edge of the profile and have
an intrinsic width of $\sim$1 ms. The cumulative probability
distribution of their intensities is best described by a power-law.
If the power-law continues to higher intensities, then
$3.7\times 10^6$ rotations are required to obtain a pulse with 20$\times$
the mean pulse flux. This number is similar to the giant pulse
rate in PSR B1937+21 and PSR B1821--24 but significantly higher
than that for the Crab.
\end{abstract}


\section{Introduction}
Despite three decades of intensive study, the mechanism producing 
pulsar radio emission is poorly understood. Fluctuations in the intensity
of the radio radiation provide important constraints on plausible mechanisms.
Single-pulse studies of bright pulsars detect a variety of
patterns in the intrinsic intensity fluctuations, including nulling
and drifting phenomena.  The distribution of integrated pulse energies, 
however, has only a modest dispersion. Johnston et al. (2001)\nocite{jvkb01}
showed that in the Vela pulsar, 99.5\% of all pulses are within
a factor of 3 of the mean flux density, $\langle S \rangle$ and
that the histogram of pulse energies is a Gaussian when plotted 
in the log. This distribution seems typical for most pulsars
\cite{hw74,rit76}.

In contrast, the Crab pulsar emits pulses with flux densities
$> 20 \times \langle S \rangle$, extending up to $> 2 \times 10^3
\langle S \rangle$ \cite{lcu+95} which were instrumental in 
the original detection of the Crab \cite{sr68}.
These giant pulses are typically broadband
\cite{mof97,sbh+99}
and of short duration, with widths of order a few $\mu$s and structure 
down to 10~ns \cite{han96b}. They are localized to the main and 
interpulse phase windows and follow an intensity distribution best 
characterized as a power law with index $\sim 3-3.5$.

The discovery of similar pulses from the millisecond pulsar
PSR B1937+21 \cite{sb95,cstt96} was surprising.
The pulses are extremely short ($\tau < 0.3 \mu$s) events
confined to small phase windows trailing the main pulse and interpulse,
again with an approximately power-law distribution of pulse energies
\cite{kt00}. Since PSR B1937+21 is the only known
radio pulsar with an estimated magnetic field at the 
light cylinder larger than that of the Crab, it has been suggested 
that this is a key parameter controlling the giant pulse phenomenon
\cite{cstt96}.

Johnston et al. (2001)\nocite{jvkb01} have recently found that 
a small subset of pulse phases from the Vela pulsar
have a very wide distribution of peak fluxes.
The pulses are localized to a phase
window $\sim$1 ms prior to the bulk of the integrated pulse emission,
are of short duration and are highly polarized. Johnston et al. (2001)
called these giant micro-pulses. It is not
clear if these events are related to true giant pulses; the largest
Vela pulses observed to date have $ S < 10\langle S \rangle$, but
these narrow pulses have peak fluxes exceeding $40\times$ the integrated peak
intensity. Kramer, Johnston \& Van Straten (2001)\nocite{kjv01} have shown that 
these giant micro-pulses have a power-law distribution and the
extended tail of the distribution 
may continue into the true giant pulse regime.
Cairns, Johnston \& Das (2001)\nocite{cjd01}, in contrast, recently 
showed that a log-normal
distribution provided an excellent fit to the flux densities in
individual phase bins across the main peak of the Vela profile.
It seems likely that the same distribution is also applicable to
other pulsars. Therefore a potential discriminator of giant pulse
activity is the change from a log-normal distribution to a power-law one.

To explore the connection between the giant pulses
in the Crab and PSR B1937+21
and the large individual pulses in the Vela
pulsar, we obtained fast time-sampled data for several
young and millisecond pulsars.
In an earlier paper \cite{rj01}
we reported the detection of giant pulses from the millisecond pulsar
with the next highest known light cylinder field, PSR B1821$-$24.
In this paper we report on our survey of young pulsars focussing
on the results obtained for PSRs B1046--58 and B1706--44.

\section{Observations and Data Analysis}
\begin{table}
\caption{Pulsars observed in the survey}
\begin{tabular}{lrrrrr}
\\
\hline\
Name & P    & DM            & Age   & Tobs & Npuls \\
     & (ms) & (cm$^{-3}$pc) & (kyr) & (hr) \\
\hline
B1046$-$58   & 123.7 & 129.1 & 20.3 & 3 &  81785 \\
J1105$-$6107 &  63.2 & 270.6 & 63.4 & 3 & 168099 \\
J1420$-$6048 &  68.2 & 359.8 & 13.0 & 4 & 205222 \\
B1509$-$58   & 150.7 & 253.2 &  1.6 & 3 &  61609 \\
J1617$-$5055 &  69.4 & 467.0 &  8.0 & 3 & 153536 \\
B1706$-$44   & 102.4 &  75.7 & 17.5 & 3 &  95536 \\
\hline
\end{tabular}
\end{table}

Table 1 lists the relevant parameters for the pulsars observed in
this survey. The pulsars were chosen on the criteria of
small spin periods and/or small age coupled with a reasonably
high flux density at 1500 MHz. The link between high energy emission
and giant pulses seems a strong one \cite{rj01}, hence we also
chose pulsars which have been detected at high energies.
PSRs B1046--58, B1706--44 and J1105--6107 are not known to 
pulse in the X-ray and their X-ray emission is likely dominated by
a surrounding nebula \cite{pkg00,bbt95,gk98}.
The first two are are pulsed in $\gamma$-rays \cite{klm+00,tab+92}, the
latter is a $\gamma$-ray detection but a search for pulses has as
yet proved unsuccesful \cite{klm+00}.
PSR B1509--58 is pulsed at both X-ray and $\gamma$-ray
wavelengths \cite{umw+93,rjm+98}.
PSRs J1420--6048 and J1617--5055 are pulsed in the X-ray and 
may also be associated with $\gamma$-ray sources
\cite{rrj01,gph97}.

Observations were made on 20-22 May 2001, with the Parkes 64-m 
radio telescope. We used the center beam of the 21-cm multi-beam system 
at an observing frequency of 1517.5 MHz. The receiver has a system equivalent 
flux density of 30 Jy on cold sky. The back-end consisted of a filterbank 
system containing 512 channels per polarization each of width 0.5 MHz for 
a total bandwidth of 256 MHz.  The polarization pairs are summed, each 
output is then sampled at 80 $\mu$s, one-bit digitised,
and written to DLT for off-line analysis.

The data were then de-dispersed at the pulsar's nominal dispersion 
measure (DM). The mean and rms of groups of 8192 samples (0.65~s) were examined 
and those which showed obvious signs of interference were discarded.
The data could then be folded synchronously with the pulsar's topocentric 
period to produce a pulse profile.  Our nominal 5-$\sigma$ sensitivity in 
80 $\mu$s is 1.3 Jy. In practice, even after removal of
high sigma points (clipping), we experienced 
substantially larger background fluctations and the rms exceeded our 
expected rms by a factor of 1.5-3. 

To confirm that we could detect conventional
giant pulses, we made short
observations of the Crab pulsar, during which giant pulses were 
detected with high significance.
Short integrations with the Vela pulsar off-axis, producing
an effective continuum flux of only 0.6 mJy, confirmed that we could detect 
the largest individual Vela-type giant micro-pulses at faint 
continuum flux levels. Finally, we observed PSR B1937+21, obtaining 
1784~s of integration ($1.15 \times 10^6$ pulses). Our sampling provides
only 19 bins across the pulse profile, but we clearly detect the giant pulse
distributions in both the main and interpulse components. In both components
the giant pulse peaks occur $\sim 1$ sample after the corresponding peak of 
the integrated pulse profile. The largest main component giant pulse obtained
had an energy 445 Jy$\mu$s, the third largest had 230 Jy$\mu$s. The phasing
and intensity distribution are well matched to the 1.4 GHz results reported
by Kinkhabwala \& Thorsett (2000)\nocite{kt00}.

\section{Results}
\begin{figure}
\centerline{\psfig{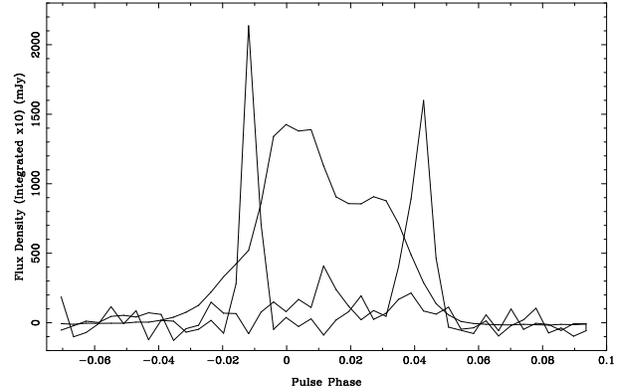}}
\caption{Two large single pulse from PSR B1046--58 shown together
with the integrated profile magnified by a factor of 10.}
\end{figure}
\begin{figure}
\centerline{\psfig{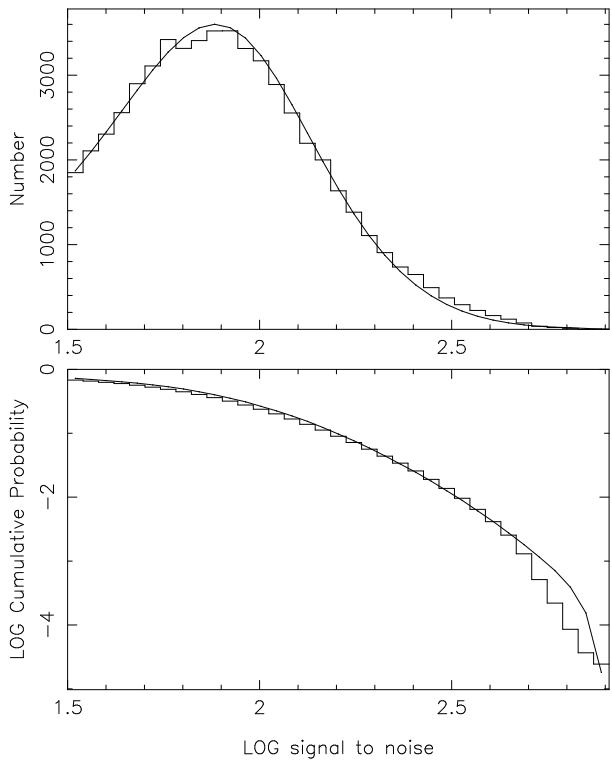}}
\caption{Top panel: Histogram of flux densities on the leading edge
of the pulse of PSR B1046--58. The solid line denotes the best fit to the data
after convolving a gaussian noise distribution of width 32.0 mJy with
a log-normal pulsar distribution with mean of 36.5 mJy and sigma (in the
log) of 0.33. Bottom panel: Cumulative probability distribution
of the data (binned) and best fit (solid line).}
\end{figure}
\subsection{PSR B1046--58}
PSR B1046--58 has a mean flux density of 5.9 mJy. It has a narrow
pulse profile  with a peak flux of 150 mJy, the highest of any of
the pulsars in our sample. We re-sampled the 80~$\mu$s data to
produce 256 phase bins across the pulsar period and examined the
$\sim$82000 single pulses. The rms per 
phase bin is 85 mJy; in $\sim$82000 pulses
we expect the largest noise sample to be 4.3$\sigma$.

On both the leading edge and trailing edges of the pulse we see 
a number of large amplitude
pulses, where the peak flux exceeds 20$\times$ the integrated flux
in these phase bins.
Examples of these pulses are shown in Figure 1.
This behaviour is reminiscent of the Vela
pulsar, where Krishnamohan \& Downs (1983)\nocite{kd83} showed
that the strong pulses were confined to the rising edge of
the profile. Kramer et al. (2001) showed that the distribution
of fluxes on the rising edge of the Vela pulsar has a log-normal
distribution with a relatively large sigma of 0.6 in the log, in
contrast to the giant micro-pulses which occur even earlier in phase
and which have a power-law distribution of fluxes.
In PSR B1046--58 the situation is less extreme. Figure 2 shows the
distribution of fluxes on the rising edge of the pulse and their 
cumulative probability distribution.
These are consistent with a log-normal fit with a sigma of 0.33 in the log.
The trailing edge of the pulse shows rather similar behaviour.
The best fit to these data is also log-normal with slightly larger
sigma of 0.45 in the log.
In contrast, the centre of the pulse is best fit with a log-normal
distribution with sigma of 0.25.

The key issue here is that log-normal distribution provides an excellent fit 
to the data at all pulse phases (Cairns et al. 2001)\nocite{cjd01}.
There is no evidence
of any power-law distribution at any phase. The variations seen
in sigma across pulse phase is consistent with the statement that the
modulation index is large towards the outside of the pulse and
small in the centre, a trend which has been known since the
mid-1970s \cite{tmh75,kd83}.

\subsection{PSR J1105--6107}
The integrated flux density of this pulsar is 1.49 mJy. The pulse
profile is double-peaked with a peak flux density of 60 mJy.
We subdivided the pulse period into 256 phase bins for a time
resolution of 250 $\mu$s and examined the $\sim$168000 pulses.
The effective rms was 100 mJy per phase bin.
Only phase bins towards the centre of the pulse had a flux density
in excess of that expected from the noise fluctuations.
The distribution of intensities in the peak flux bin was consistent with a 
log-normal distribution with sigma of 0.3 in the log convolved with
gaussian statistics for the noise. There was no
evidence of a power-law tail to the distribution in any phase bin.

\subsection{PSR J1420--6048}
The pulsar was observed for a total of 4 hr, split into two separate 
observations.
The integrated flux density for both observations is only 0.95 mJy.
The pulsar has large
duty cycle; as a result its peak flux is only 12 mJy. No large samples
were detected at any pulse phase. Subdividing the pulse into
256 phase bins and examining the single pulses yielded statistics which
were consistent with the noise level of 90 mJy. One pulse out of $\sim$210000
had, however, a peak flux of 800 mJy, this is the only sample in the entire
data set with a flux greater than 5 times the noise rms (450 mJy).
This high point is located on the leading edge of the second peak, 
where the integrated flux density is only 2.1 mJy.
However, re-examination of individual 80 $\mu$s samples, revealed that
this high flux is entirely due to one sample.  As the DM
of the pulsar ensures that any real signal should be 
smeared by $\sim$7 samples, we therefore
believe this does not originate from the pulsar but is likely to
be interference or an instrumental glitch.

\subsection{PSR B1509--58}
The integrated flux density during our 3 hr observation was only
0.96 mJy. This pulsar has a broad single component with a peak flux density
of only 8 mJy. The DM is large, effectively smearing our
time resolution to 400 $\mu$s. No large samples were detected during
the observation. We then subdivided the pulse period into 256 phase bins
for a time resolution of 590 $\mu$s
and examined the $\sim$65000 single pulses. The effective rms was 70 mJy per
phase bin, no single phase bin had a flux density in excess of
5$\sigma$.
\begin{figure}
\centerline{\psfig{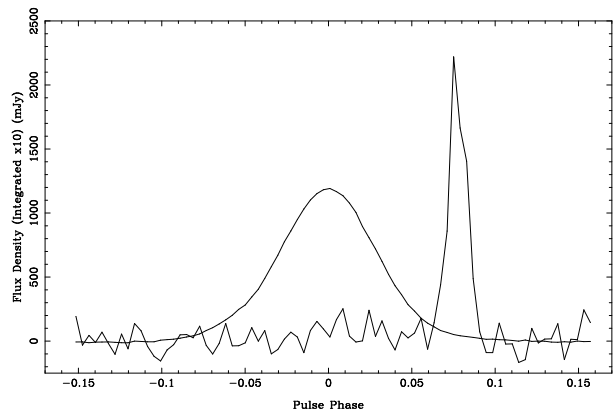}}
\caption{A large single pulse from PSR B1706--44 shown together
with the integrated profile magnified by a factor of 10.
Virtually all the large pulses are located at this pulse phase.}
\end{figure}

\subsection{PSR J1617--5055}
The pulsar is very weak with a flux density of only 0.15 mJy in our
3 hour observation. The pulsar is significantly scatter-broadened
at this frequency and the peak flux is only 1.5 mJy. We subdivided
the pulse period into 128 phase bins and examined each of
the $\sim$150000 pulses. No single phase bin of any pulse was in
excess of 400 mJy or 5$\times$ the noise level.
This is consistent with Kaspi et al. (1998)\nocite{kcm+98} who also 
reported a lack of giant pulses from this pulsar.
\begin{figure*}
\centerline{\psfig{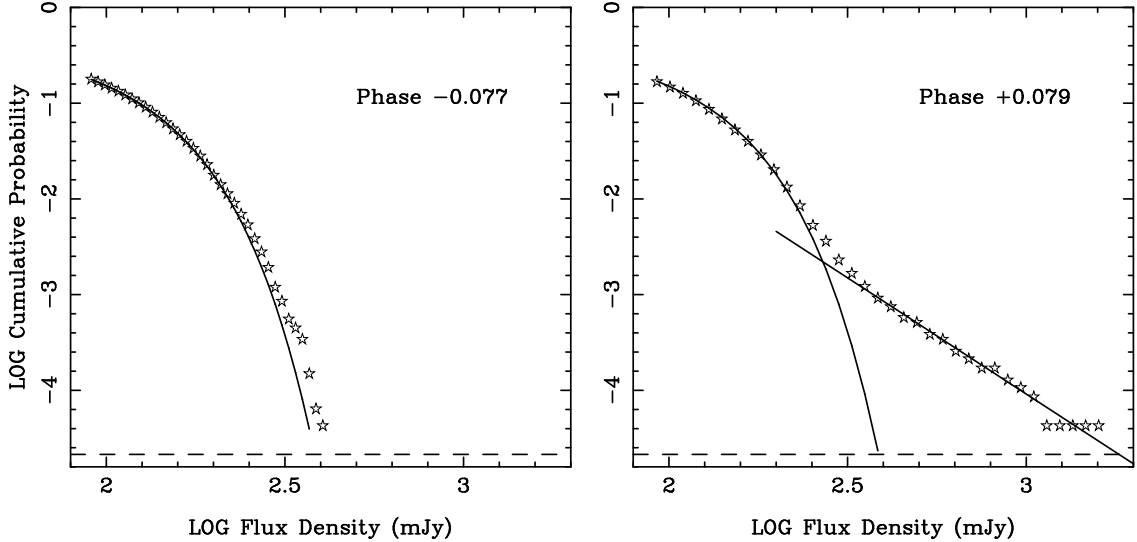}}
\caption{Cumulative probability distributions for two phase bins.
The solid line is the expected distribution from Gaussian noise with an 
rms of 90 mJy. The right panel shows an extra solid line representing
a power-law of slope --2.7.
The dashed line is the reciprocal of the number of pulses.}
\end{figure*}

\subsection{PSR B1706--44}
This pulsar was observed twice, each observation was 1.5 hr in length.
The integrated flux densities were 8.63 and 7.70 mJy in the two observations.
This pulsar has a profile which consists of a single component 
with a peak flux of $\sim$100 mJy.
Analysis of each 80$\mu$s sample revealed that many strong samples
were detected in a small phase window located far into the wings
of the trailing edge of the profile. No such samples were detected on
the leading edge and only a few at the peak of the pulse profile.
It was clear that the large pulses were resolved on this timescale.
We then divided the pulse period into 256 phase bins and the $\sim$46000
single pulses from each observation were examined. The rms per
bin is 92 mJy.
Figure 3 shows the pulse with the largest peak flux density. The flux
density in this phase bin is $500\times$ the flux density in the
integrated profile at this phase, but the integrated flux is only
4 times that of the integrated profile (i.e. it would not be classified
as a true giant pulse). The pulse is clearly resolved
with a half width of $\sim$1 ms. This width must be intrinsic; the
DM smearing is only 120 $\mu$s and the scattering time at this frequency
is negligible \cite{jnk98}. Of the $\sim$93000 pulses collected,
52 had a peak flux density in excess of $50\times$ the integrated flux
density at a given phase. All these pulses are located between
phases 0.06 and 0.095.

Figure 4b shows the cumulative probability distribution of the
intensities at phase 0.079. The mean flux density at this phase
is 4 mJy, therefore only pulses with intensities greater than
$80\times$ this can be detected at the 3-$\sigma$ level.
The distribution clearly deviates from that expected from gaussian noise
and at high flux levels can best be described with a power law with
index --2.7. This power law index is very similar to that seen in
the Crab pulsar and PSR B1937+21. In contrast Figure 4a shows intensities 
for a bin on the
rising edge of the profile with the same integrated flux density
as for phase 0.079. No large intensities are seen and the
distribution is consistent with noise.

There are no true giant pulses seen in our data set for PSR B1706--44.
The maximum flux density of any single pulse was 
only $\sim$4$\times$ the mean integrated flux
density. Figure 5 shows the distribution of fluxes summing over
all phase bins. The fluxes have been normalised to take into account
the effects of interstellar scintillation. The width of the
distribution is essentially determined by the receiver noise.
However, there is an excess of counts at high fluxes - there
are 782 pulses with flux densities greater than 2.4$\times$ the mean
flux density but only 490 pulses with flux densities less than -0.4$\times$
the mean. This indicates the pulsar flux is
likely log-normally distributed.
\begin{figure}
\centerline{\psfig{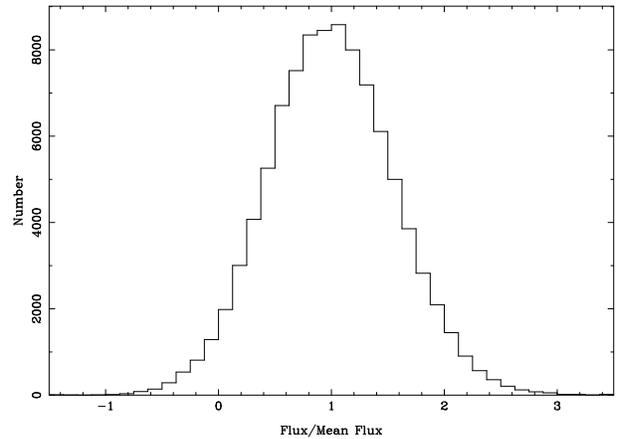}}
\caption{Distribution of the integrated flux of all 96000 pulses
of PSR B1706--44 normalised by the mean flux (computed on a time-scale
significantly shorter than the scintillation timescale) to remove
the effects of interstellar scintillation.}
\end{figure}

It is our belief that the power-law distribution shown
in Figure 3b is indicative of giant pulse behaviour.
If this power law continues to
larger fluxes, then $3.7\times 10^6$ rotations are needed before
reaching a pulse with 20 times the mean flux density. This number of
rotations is within a factor of 10 of the giant pulse rate in both
PSR B1937+21 and B1821--24 but significantly higher than that
of the Crab.

\section{Discussion}
In the Vela pulsar, log-normal statistics are adequate to fit the
distribution of flux across the bulk of the pulse profile
(Kramer et al. 2001, Cairns et al. 2001)\nocite{kjv01,cjd01}
but the width of the distribution is significantly
larger at the edge of the profile than in the middle.
In addition to this, there are giant micro-pulses which occur well
before the main pulse phase. These giant micro-pulses have 
a half-width of $\sim$200 $\mu$s, and are not at a fixed phase,
but have an inherent `jitter' of about 1 ms \cite{jvkb01}.
Their distribution is best described by a power-law
(Kramer et al. 2001)\nocite{kjv01}.
We have found new examples of both these phenomena in our
current survey.

PSR B1706--44 shows an additional example of giant micro-pulses.
This time, however, the giant micro-pulses are
located on the trailing edge of the pulse, and are somewhat wider
then in Vela with a half-width close to 1 ms. Again there is some phase
jitter as to the location of the pulse maximum. For this pulsar also,
the distribution of fluxes is clearly power-law at high amplitudes.

In PSR B1046--58 we clearly see large amplitude pulses on
both the leading and trailing edges of the integrated pulse profile.
The intensity distribution at these phases, however, is adequately
described by a log-normal distribution with moderate width.
The peak fluxes achieved are in excess of 20$\times$ the mean flux density
but there is no evidence for a power-law tail to
the distribution. In this regard
they are similar to the fluctuations seen on the rising edge of
the Vela pulsar.

No giant micro-pulses are detectable in PSRs J1105--6107,
J1420--6048, B1509--58 or J1617--5055. However, the sensitivity to 
giant micro-pulses in these pulsars is not as good as for PSR B1706--44 or
B1046--58. To demonstrate this, let's assume that any 
giant pulses would be similar
to those seen in PSR B1706--44; i.e. they would have an intrinsic 
width of 0.5 ms, a power-law index of --3.0 and occur every $10^7$ rotations.
The brightest giant pulse in the observing time for PSRs J1105--6107,
J1420--6048, B1509--58 and J1617--5055 would therefore be
less than 500 mJy (in 0.5 ms) and not detectable in the noise.
Giant pulses could only be detected in these pulsars if they
were intrinsically very narrow, and/or occured much more frequently
than in either Vela or PSR B1706--44.

We find, as in previous studies (e.g. Taylor et al. 1975\nocite{tmh75}),
that the modulation of pulse intensity 
is larger in the wings of the profile of Vela-like pulsars.
PSR B1046--58 is a good example of this behaviour: its phase resolved
intensity distributions are well described by a log-normal distribution
whose width increases towards the profile edge.
The giant micro-pulses in Vela 
and now PSR B1706--44 appear to represent a distinct pulse population, with
emission phases well separated from the bulk of the integrated
pulse profile. Both of these phenomena represent enhanced intensity 
fluctuations.

Observations of the Crab pulsar at infra-red, optical and higher energies 
(e.g. Lundgren et al. 1995, Patt et al. 1999, Romani et al. 2001)
\nocite{lcu+95,puz+99,rmc+01} show that there are no detectable 
intensity fluctuations in the (incoherent) high energy emission associated 
with the radio giant pulses. Thus we must conclude
that giant pulses represent a variation in the coherence of 
the particle distribution. An important question is why coherence 
fluctuations appear to be strongest at the pulse edges.

One clue may be found in the connection of pulse width to altitude. 
Crudely, in a dipole polar cap of radius $r$, the width of the edge of the open 
zone maps as $\delta \phi \propto r^{1/2}$. 
If interpreted as the edge of a polar zone, the leading and trailing 
edges of the profiles where the modulation index increases represent
locations 1.5-2$\times$ higher than that of the 
mean pulsar emission. If we adopt the classical polar gap picture, with a 
pair formation front producing a dense plasma at a few stellar radii, 
then we can identify pulse intensity fluctuations with the 
growth of instabilities in the outflowing pair wind. Larger widths
represent higher altitudes, and in this picture larger phase-space amplitudes 
for the instability-driven fluctuations, which have 
had more time to grow in the outflow. Alternative
pictures exist, such as the slot-gap scenario of Arons (1983)\nocite{aro83b}
for which the gap height is larger at the edge of the polar cap.
Again there could be a plausible connection with increased instability.

In both PSR B1706--44 and Vela the giant micro-pulses are located
far from the peak of the pulse profile, representing an emission
altitude $\sim$4$\times$ higher. It is plausible that
instability growth reaches a different regime at these heights. Clearly a 
true power-law intensity distribution suggests that the instabilities 
have left the (linear) stochastic growth regime (Cairns et al. 2001).
Power-law behaviour suggests a causal connection between 
different intensity scales such as a cascade from 
one energy scale to another, or a self-similar coupling
as in a saturated growth scheme. Such processes might 
develop late (higher) in the polar wind outflow.

The giant micro-pulse emission is however well separated from the bulk
of the radio emission in these objects, suggesting 
instead an independent origin.
There is good evidence that for the Crab and possibly PSR B1821--24,
the giant pulse radio emission is co-located with high energy emission 
in the outer magnetosphere, where strong power law X-ray components indicate 
sites of dense pair production. Vela and PSR B1706-44 are also
$\gamma$-ray emitters; in the outer magnetosphere picture
this high energy emission comes from the pole opposite to that viewed 
in the radio. They do not show strong X-ray pulses and so it is 
perhaps not surprising that we have not found giant pulse emission 
coincident with the $\gamma$-rays in these objects, as the Earth line-of-sight
evidently does not sample regions of dense pair production.
Nevertheless, there may be a high energy connection to the
giant micro-pulse emission if the outer magnetosphere above the radio pole is
also actively producing pairs. The high energy emission from beyond 
the null charge surface would not be visible from this pole, but some 
pairs from the gap would be expected to flow inward past the null 
charge surface. We can speculate that this plasma mirrors in 
the converging field lines above the radio pole and that this mirrored,
counter-streaming population would suffer instability growth and
produce the giant micro-pulse components.

A final question concerns the asymmetry of the giant 
micro-pulse component. Why does it lead in the case of Vela, but 
lag for PSR B1706$-$44? For Vela, the radio pulse has a steep rise and
slow fall suggesting the leading edge of a cone; the
giant micro-pulse component is at higher altitudes on the same side. For
PSR B1706$-$44, one would then infer that the main radio pulse is the 
trailing edge of a cone, although this is morphologically less clear.
If there is a high energy connection, then it is intruiging to note that 
the leading $\gamma$-ray pulse is stronger for Vela, while for PSR B1706$-$44,
the trailing component is stronger at most high energies.
More examples are clearly needed to see if the dominance of 
leading versus trailing giant micro-pulses has a deterministic connection 
with other pulsar emission and if some global asymmetry in the 
magnetosphere geometry controls this choice.

\section*{Conclusions}
We have found evidence for giant micro-pulses in PSR B1706--44 which
are very similar to those seen in the Vela pulsar. They are located
in a small window of pulse phase, have a duration of $\sim$1 ms and
significant phase jitter. Their amplitude distribution is power-law
and so may extend into true `giant pulses' if observed for long enough.
It is unclear whether the giant micro-pulses are just another 
manifestation of pulsar `weather' associated with the standard radio
emission from pulsars or whether they are more closely related to
high-energy phenomena and the classical giant pulses.

No giant pulses or any other giant micro-pulses were detected 
within our sensitivity limits in any of the other 5 pulsars in our survey.
Where it was possible to measure amplitude distributions, these
were generally log-normal as surmised by Cairns et al. (2001).

\section*{Acknowledgments}
The Parkes telescope is funded by the Commonwealth of Australia
for operation as a National Facility managed by CSIRO. 
The 512 channel filterbank system used in the observations
was designed and built at the Jodrell Bank Observatory, University 
of Manchester.  Observing software was provided by the Pulsar 
Multibeam Survey group.

\bibliographystyle{mn}
\bibliography{modrefs,psrrefs,crossrefs}

\end{document}